\documentclass[12pt]{article}
\usepackage{hyperref}
\usepackage{cite}
\usepackage{color}
\usepackage{graphicx}
\usepackage{amsmath}
\usepackage{amssymb}
\usepackage{xspace}

\makeatletter
\@addtoreset{equation}{section}

\makeatletter
\renewcommand\section{\@startsection {section}{1}{\z@}%
                                   {-3.5ex \@plus -1ex \@minus -.2ex}
                                   {2.3ex \@plus.2ex}%
                                   {\normalfont\large\bfseries}}
\renewcommand\subsection{\@startsection{subsection}{2}{\z@}%
                                     {-3.25ex\@plus -1ex \@minus -.2ex}%
                                     {1.5ex \@plus .2ex}%
                                     {\normalfont\bfseries}}

\def\baselinestretch{1.2}
\newcommand{\be}{\begin{equation}}
\newcommand{\ee}{\end{equation}}
\newcommand{\beq}{\begin{eqnarray}}
\newcommand{\eeq}{\end{eqnarray}}

\newcommand{\gone}[1]{{}}

\begin{document}
\begin{titlepage}
\begin{flushright}
MAD-TH-15-06
\end{flushright}

\vfil

\begin{center}

{\bf \Large
Dynamics of ${\cal N}=4$ supersymmetric field theories in 2+1 dimensions and their gravity dual
}

\vfil

William Cottrell, James Hanson, and Akikazu Hashimoto

\vfil

Department of Physics, University of Wisconsin, Madison, WI 53706, USA

\vfil

\end{center}

\begin{abstract}
\noindent In this note we consider ${\cal N}=4$ SYM theories in 2+1
dimensions with gauge group $U(N)\times U(M)$ and $k$ hypermultiplets
charged under the $U(N)$.  When $k > 2(N-M)$, the theory flows to a
superconformal fixed point in the IR. Theories with $k <2(N-M)$, on
the other hand, flows to strong coupling. We explore these theories
from the perspective of gravity dual. We find that the gravity duals
of theories with $k < (N-M)$ contain enhancons even in situations
where repulson singularities are absent. We argue that supergravity
description is unreliable in the region near these enhancon points.
Instead, we show how to construct reliable sugra duals to particular
points on the Coulomb branch where the enhancon is screened. We
explore how these singularities reappear as one moves around in
Coulomb branch and comment on possible field theory interpretation of
this phenomenon. In analyzing gauge/gravity duality for these models,
we encountered one unexpected surprise, that the condition for the
supergravity solution to be reliable and supersymmetric is somewhat
weaker than the expectation from field theory. We also discuss similar
issues for theories with $k=0$.
\end{abstract}
\vspace{0.5in}

\end{titlepage}
\renewcommand{\baselinestretch}{1.05}  

\section{Introduction}

Supersymmetric field theories exhibit rich dynamical phenomena which
are nonetheless susceptible to explicit analysis. The nature of the
dynamics vary depending on the number of dimensions and the amount of
supersymmetries.  Naturally, the most studied class is field theories
in 3+1 dimensions. The ${\cal N}=4$ theories are conformal. ${\cal
  N}=2$ theories have quantum corrected moduli space. ${\cal N}=1$
theories generically exhibit dynamically generated superpotentials
that give rise to vacuum selection and symmetry breakings. Quite a bit
is also known about supersymmetric field theories in 2+1
dimensions. Certain aspects of dynamics in 2+1 dimensions can be
inferred by looking at the system as a theory in 3+1 dimensions
compactified on a circle. The basic picture in the case with 8
supercharges was considered in \cite{Seiberg:1996nz}.  More recently,
the case with 4 supercharges was studied in
\cite{Aharony:2013dha,Aharony:2013kma}.

A phenomena that is unique to 2+1 dimensions is dynamical breaking of
supersymmetry in Chern-Simons-Yang-Mills theories with ${\cal N}=1$
\cite{Witten:1999ds}, ${\cal N}=2$, or ${\cal N}=3$
\cite{Bergman:1999na,Ohta:1999iv} supersymmetries. The spontaneous
breaking of supersymmetry for these models was argued based on
computation of the Witten index and the $s$-rule applied to their
brane construction.  When spontaneous breaking of supersymmetry occurs
in a weakly coupled theory, one can analyze the vacuum energy,
condensate, and the presence of Goldstone fermions explicitly
(although somewhat messily.) From dimensional considerations and
counting of parameters, one expects the vacuum energy to scale, for
$(N/k) \gg 1$ as
\be {\cal E} = \# (N/k)^\# g_{YM}^2 \ee
where $g_{YM}^2$ in $2+1$ dimensions has the dimension of energy, and
$\#$'s are dimensionless constants of order one that should be
computable from first principles.\footnote{This relation will be
  generalized when the model is generalized such as including large
  number of flavors.}  Such analysis is not immediately possible for
these models because they are strongly coupled.  Attempts to analyze
these features often involve modifying the theory in the ultra-violet
and tuning the parameters so that DSB takes place in a weakly coupled
regime, e.g.\ \cite{Suyama:2012kr}.

One possible approach to access these features in this model is to
invoke gauge-gravity duality where one hopes to capture the relevant
supersymetry breaking dynamics in terms of degrees of freedom that are
weakly coupled in the gravity description. This program has met with
limited success so far \cite{Giecold:2013pza,Cottrell:2013asa}, in
that the dual supergravity background contains singularities making
its effective dynamics beyond the scope of the supergravity
description. Perhaps one can work harder at extracting meaningful
effective dynamics along the lines of \cite{Michel:2014lva}, but how
precisely to do that for our purpose is not completely clear at the
moment.

The goal of this article is to retreat to a simpler system where the
field theory dynamics is under better control and to explore the
singularities which arise in the gravity dual.  Specifically, we
consider a class of supersymmetric field theories in 2+1 dimensions
with ${\cal N}=4$ supersymmetries. These models generically have a
moduli space of vacua, with various branches, whose structure can be
subject to quantum corrections.  Some points in moduli space such as
the point where two branches meet often plays a special role. Presumably,
the full diversity of phenomena on the field theory side is reflected
on the gravity side in the resolution of singularities.  It is
tempting to propose that mapping out such correspondences would
eventually have profound impact on understanding black hole,
cosmology, and other gravitational phenomena involving singularities.

\section{${\cal N}=4$ field theories in 2+1 dimensions and their supergravity dual\label{sugrasec}}

In this section, we will review the supergravity solution which will
be the focus of our analysis. The background in question was
constructed explicitly in \cite{Aharony:2009fc} where much of the
details and the conventions can be found.\footnote{See also
  \cite{Bertolini:2000dk} for earlier construction of supergravity
  background with fractional branes.} Here, we will summarize key
features in order to make this paper self contained, but the readers
are referred to \cite{Aharony:2009fc} for a more thorough account.

\subsection{Basic Setup}

\begin{figure}
\centerline{\begin{tabular}{cc}
\raisebox{5ex}{\includegraphics[width=2.5in]{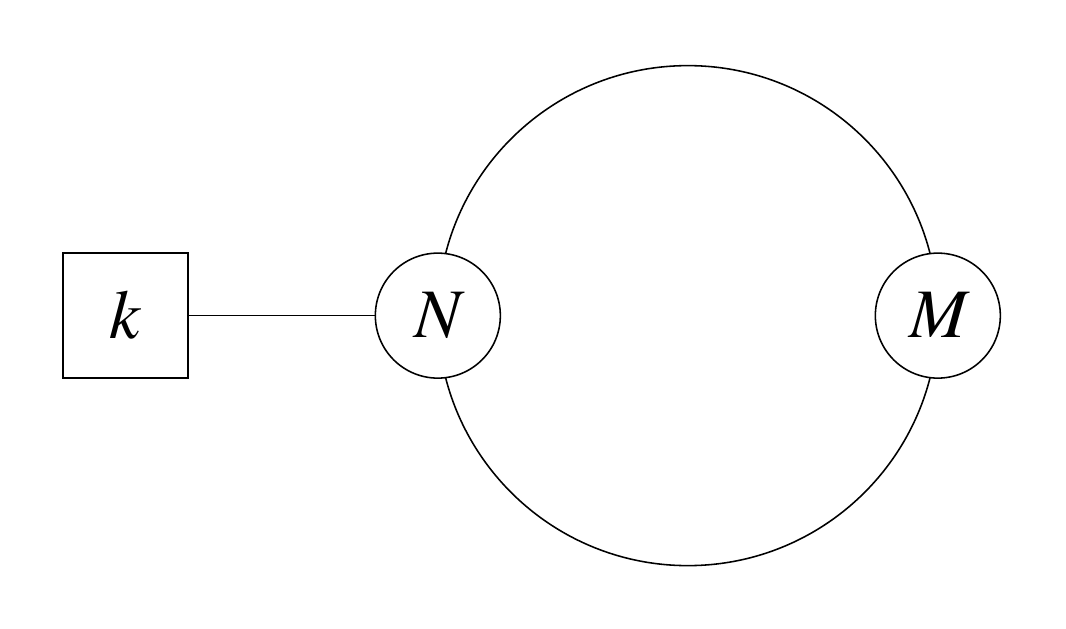}} & \includegraphics{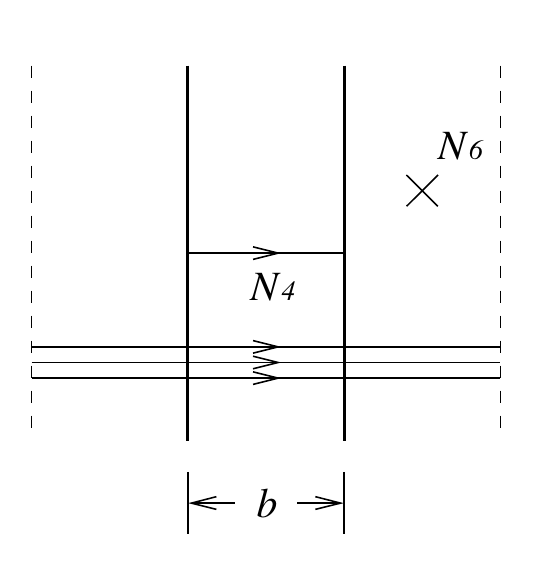} \\
(a) & (b)
\end{tabular}}
\caption{\label{figa} Quiver diagram (a) and Hanany-Witten diagram (b) of $U(N)\times U(M)$ with $k$ flavors.}
\end{figure}

The class of theories we consider consists of (2+1)d ${\cal N}=4$ SYM with 
gauge group $U(N)\times U(M)$ and $k$ fundamental hypermultiplets 
charged under $U(N)$.   They are represented by a
circular quiver of the form illustrated in figure \ref{figa}.a. Such a
model can be constructed from the type IIB brane configuration illustrated in
figure \ref{figa}.b. The construction involves 2 NS5-branes and $k$
D5-branes, $N_2 = N$ ``integer'' D3-branes winding all the way around
the $S^1$ of period $L$, and $N_4 = M-N$ ``fractional'' D3-branes
suspended between the two NS5-branes separated by the distance
$bL$. In the $\alpha' \rightarrow 0$ zero slope limit, most of the
string states decouple and we obtain a 3+1 dimensional defect theory
on $R^{1,2} \times S^1$. In the limit that $L$ goes to zero, momentum
modes along the $S^1$ decouples and we obtain a theory in 2+1
dimensions.

\subsection{Supergravity solution}

The gravity dual is most easily constructed by T-dualizing along $S^1$ which maps the 2 NS5-branes to $TN_{2}$  (which approaches the $C^2/Z_2$ ALE geometry in the $L\rightarrow 0$ limit), D5-branes  to D6-branes, integer D3-branes to D2-branes, and fractional D3-branes to fractional D2-branes, which are D4-branes wrapping the collapsed 2-cycle at the tip of the $C^2/Z_2$ ALE.

One can then think of the IIA solution as a dimensional reduction of
M-theory on $R^{1,2} \times (C^2/Z_2) \times TN_k$ to which we add the back reaction of D2 and D4 branes sources. It is therefore natural to consider an ansatz where $R^{1,2} \times (C^2/Z_2) \times TN_k$ gets warped as a result of fluxes sourced by the D2 and the D4-branes. 
 
The ansatz considered in  \cite{Aharony:2009fc} is
\beq ds^2 &=& H^{-2/3} (-dt^2+dx_1^2+dx_2^2) + H^{1/3} (ds_{ALE}^2+ds_{TN_k}^2), \\
G_4 &=& dC_3 =  dt \wedge dx_1 \wedge dx_2 \wedge d H^{-1} + G_4^{SD}, \label{g4}\\
G_4^{SD} &=& d(l V \omega_2 \wedge \sigma_3 + 2 \alpha \omega_2 \wedge d \psi) \eeq
Let us make few comments regarding this ansatz.

\begin{itemize}

\item The Taub-NUT metric is given by
\be ds_{TN_k}^2 = V(r)^{-1} (d r^2 + r^2 (d \theta^2 + \sin^2\theta d \phi^2)) +V(r)
R_{11}^2 k^2
\left( d \psi - {1 \over 2}\cos \theta d \phi \right)^2 \label{TNmetric} \ee
with
\be V(r) \equiv
\left(1 + {k R_{11} \over 2 r} \right)^{-1}, \qquad R_{11} = g_s l_s\ , \ee
for the range of coordinates\footnote{The case $k=0$ will require some modifications.} $0 \le r < \infty$, $0 \le \theta \le
\pi$, $0 \le \phi \le 2 \pi$, $0 \le \psi \le 2 \pi/k$.  The parameter $k$  is related to the number of D6-branes
\be k = N_6 \label{kn6} \ . \ee

\item 1-form $\sigma_3$ lives in the Taub-NUT space

\be {1 \over 2} \sigma_3 \equiv d \psi - {1 \over 2} \cos \theta d \phi \ . \ee

\item The 2-form $\omega_2$ is  dual to the collapsed 2-cycle of the $C^2/Z_2$
ALE. It is normalized so that
\be \int_{ALE} \omega_2 \wedge \omega_2 = {1 \over 2} \ . \ee

\item The parameter $l$ parameterizes the magnitude of $G_4^{SD}$. The
seemingly trivial parameter $\alpha$ which does not contribute to
$G_4^{SD}$ on the account of $\omega_2$ and $d \psi$ being closed,
will turn out to be important for quantizing charges. 

\item To solve the M-theory equation of motion, the warp factor $H$ must satisfy the Poisson equation
\be 0 =  \left(\nabla^2_y+ \nabla^2_{TN} \right) H + {l^2 V^4\over 2  r^4}  \delta^4(\vec y)
+ (2 \pi l_p)^6  Q_2 \delta^4(\vec y) \delta^4(\vec r)
\label{harm} \ , \ee
where $\vec y$ is a four vector parameterizing $C^2/Z^2$, and $\vec r$ is a four vector parameterizing the Taub-NUT space. We have introduced another parameter $Q_2$ which corresponds to the magnitude of the D2-brane source which we will describe in more detail below.

\item When expressed in terms of IIA supergravity fields, the solution takes the form
\beq ds_{IIA}^2 &=& H^{-1/2} V^{1/2} (-dt^2 + dx_1^2+dx_2^2) + H^{1/2} V^{1/2} ds_{ALE}^2
\label{iiaf} \cr
&& \qquad  + H^{1/2} V^{-1/2} (dr^2 + r^2 (d \theta^2 + \sin^2 \theta d \phi^2)),\\
A_1 & = & -{1 \over 2} R_{11}k  \cos \theta d \phi, \\
A_3 & = & -(H^{-1}-1) dt \wedge dx_1 \wedge dx_2 - l V \omega_2 \wedge \cos \theta d
\phi,\\
B_2 & = &   -{2 \over R_{11} k} (l V \omega_2 +  \alpha \omega_2),  \\
e^{\phi} &=& g_s  H^{1/4} V^{3/4}. \label{iial}
\eeq
It is convenient to introduce a field variable $b$ by the relation
\be B_2 = (2 \pi)^2 \alpha' b\,  \omega_2, \ee
so that
\be b(r) = - {2 \over  (2 \pi l_s)^2 R_{11} k} (l V + \alpha) \ee
is dimensionless.

\item Parameters $\alpha$ and $l$ are fixed by imposing quantization of the D4-brane charge and the asymptotic behavior of $b(r)$ at $r=\infty$. 

Requiring that the D4 Page charge is integrally quantized leads to the relation\footnote{A slightly different treatment is required for the case of $k=0$.}
\be 2 \pi \alpha = (2 \pi l_s)^3 g_s N_4 \ . \label{alpha} \ee
One can then read off how $l$ depends on $b_\infty$ and $N_4$ 
\be l = - 2 \pi^2 k l_s^2 R b_\infty - \alpha  = - (2 \pi)^2 g_{YM}^2 \alpha'^2 \left(N_4 + {k b_\infty \over 2} \right) \ . \label{ell} \ee
With supergravity parameters $l$ and $\alpha$ specified in terms of field theory data $b_\infty$ and $N_4$, we can write $b(r)$ more compactly as
\be b(r) = b_\infty V(r) - {2 N_4 \over k} (1 - V(r)) \label{br} \ee
so that $b(0) = (-2 N_4/k)$  and $b(\infty) = b_\infty$.

Note that something slightly unexpected has happened. The magnitude of
gauge invariant field strength $G_4$ parameterized by $l$ depends on
$b_\infty$ and is continuous, whereas seemingly gauge dependent
parameter $\alpha$ depends on $N_4$ and is discrete. 

\item The last remaining parameter of the supergravity solution that needs to be related to the field theory data is $Q_2$ in (\ref{harm}). This parameter should be set so that the D2-page charge is integrally quantized, leading to the relation
\be Q_2 = N_2 + b_0 N_4 + {N_6 b_0^2 \over 4} = N_2 - {N_4^2 \over k} \label{Q2} \ee

\item To summarize, parameters $\alpha$, $l$, $k$, and $Q_2$ which
  appear as part of the supergravity ansatz is related to $N_2$,
  $N_4$, $N_6$, and $b_\infty$ by relations (\ref{kn6}), (\ref{alpha}), (\ref{ell}), and (\ref{Q2}). $N_2$, $N_4$, $N_6$, and $b_\infty$
  have natural interpretations on the field theory side. $N_2$, $N_4$,
  and $N_6$ take on integer values, whereas $b_\infty$ takes on
  continuous values in the range $0 \le b_\infty \le 1$.

\item With $Q_2$, $l$, and $\alpha$ parameterized in terms of $N_2$, $N_4$, and $b_\infty$, the Maxwell D2 and D4 charges becomes
\beq
Q_4^{Maxwell} &=& N_4 + {1 \over 2} b_\infty N_6, \label{qfourmax}\\
Q_2^{Maxwell} & = & N_2 + b_\infty N_4 + {1 \over 4} b_\infty^2 N_6. \label{qtwomax}\eeq

\item In order to identify the gravity solution we constructed with the brane configuration having the linking numbers illustrated in figure \ref{figa}.b and not as illustrated in figure \ref{figb}, we examine the probe action of D4-branes in this background. 

\begin{figure}
\centerline{\includegraphics{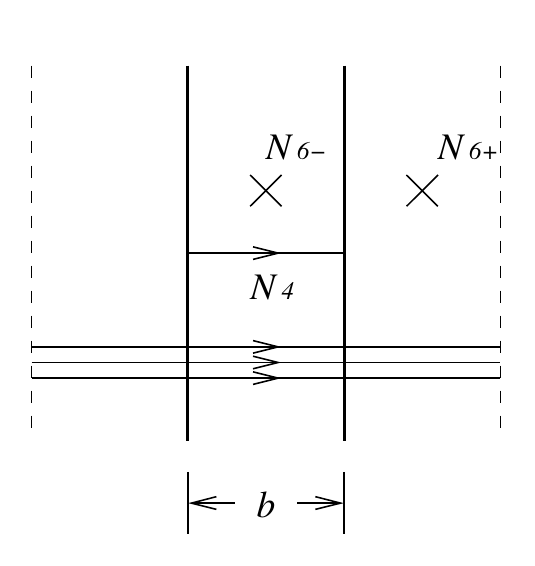}}
\caption{\label{figb} A brane configuration with flavors charged under
both gauge groups.}
\end{figure}

Consider a D4 or an anti D4 probe wrapping the collapsed 2-sphere at the tip of the ALE dual to $\omega_2$ threaded with $n$ units of magnetic flux. Provided\footnote{Here, we correct a subtle sign error in (2.47) of \cite{Aharony:2009fc}.}
\be 
(lV +\alpha)\pm 2 \pi^2 l_s^2 R_{11} k n    < 0\ , \label{bound} \ee
where $\pm$ corresponds to the D4 and the anti D4 respectively,
the potential term in the DBI and the WZ terms cancel, giving rise to leading term in the derivative expansion of the form
\beq S &=& T_4 \left( \left(2 \pi^2 l_s^2 n \pm {\alpha \over kR_{11}}\right)V^{-1}\pm {l
\over kR_{11}} \right)(\dot r^2 + r^2 \dot \theta^2 + r^2\sin^2 \theta \dot \phi^2)  \cr
& \equiv & {2 \pi^2 l_s^2 T_4 \over g_{eff}^2(r)} (\dot r^2 + r^2 \dot \theta^2 + r^2\sin^2
\theta \dot \phi^2),   \eeq
where
\be {1 \over  g_{eff}^2(r)} =  (n \pm b_\infty) + {R_{11} (nk \mp 2 N_4 ) \over 2 r} \ . \ee
Performing the standard map between gauge theory and dual string
theory parameters,
\be R_{11} = g_s l_s, \qquad r = 2 \pi l_s^2 \Phi, \qquad g_s = g_{YM}^2 (2 \pi)^{-(p-2)}
l_s^{-(p-3)}, \ee
where $\Phi$ is the vacuum expectation value of the scalar field in
along the Coulomb branch of the  ${\cal N}=4$ gauge theory, we find
\be
{1 \over  g_{eff}^2(\Phi)} = (n \pm b_\infty) + {g_{YM}^2 (nk \mp 2 N_4) \over 4 \pi \Phi} \ \label{geff}
. \ee

If one takes,  $n = 0$ for the ``+'' (D4-brane) and $n=1$ for the
``$-$'' (anti-D4-brane), the effective gauge couplings take the form
\beq
{1 \over  g_{eff1}^2(\Phi)} &=& b_\infty  - { g_{YM}^2 N_4 \over 2 \pi \Phi} \ , \label{geff1} \\
{1 \over  g_{eff2}^2(\Phi)} &=& (1 - b_\infty) + {g_{YM}^2(k + 2 N_4) \over 4 \pi \Phi} \label{geff2} \ .
\eeq

This is interpretable as the expected running of the dimensionless
coupling of the $U(N_2+N_4)$ and the $U(N_2)$ gauge groups, with $k$
fundamentals charged under $U(N_2)$.\footnote{See page (8.43)-(8.44) of
  \cite{ClossetThesis} where field theory manifestation of such
  running  is discussed.}

The dimensionful gauge coupling for $U(N_2+N_4)$ and $U(N_2)$ in 2+1
dimensions at scale $\Phi$ is given,respectively by multiplying
$g_{eff1}(\Phi)$ and $g_{eff2}(\Phi)$ by $g_{YM2})$. At the UV fixed
point, they are, respectively, $g_{YM}^2 b_\infty$ and $g_{YM}^2
(1-b_\infty)$.

\item We know from field theory calculations that the perturbative
  moduli space of this theory is given by a multi-center Taub-Nut
  geometry described in (13) of \cite{Tong:1998fa}.  This space is $4
  (N+M)$ real dimension hyper-Kahler geometry corresponding to $(N+M)$
  D3 segments in Hanany-Witten brane construction, but we can infer a
  $4$ real dimensional subspace by keeping the position of the
  $(N+M-1)$ D3 segments fixed. This can be described, treating one of
  the $N$ D3-branes of the $U(N) \times U(M)$ theory as a probe, as
  the anti D4-brane probe with one unit of $D2$ charge.  In order to infer the full hyper-Kahler structure,  
  dualize the world-volume $(2+1)$ gauge field into a periodic scalar.  In additon to 
  the gauge field kinetic term, we must also account for the Wess-Zumino term:
\be
S \sim \left(B_{2}+2\pi \alpha' f_2 \right) \wedge C_{3}+... \\
\ee
where $f_2$ is the field strength of the world volume gauge field. 
We have dropped terms which end up not contributing to integral over the two cycle
wrapped by the probe. To construct the dual scalar form of the action, we add a lagrange 
multiplier to enforce the constraint $df_2=0$ and then treat $f_2$ as a 
free field.  The relevant terms in the lagrangian are then:
\beq
S_{\overline{D4}}& \sim &\int \left(\tilde{V} |f_2|^2 -\frac{1}{4\pi}\left( d\varphi-(k+2N_{4})\cos\theta d\phi\right) \wedge f_2+...\right) 
\eeq
where we have defined
\be
\tilde{V} \equiv g^{-2}_{eff2}\ . \ee
As usual, the dual scalar $\varphi$ is compact with periodicity $4 \pi$.
Now, integrate out $f_2$, and one gets
\be
S_{\overline{D4}} = \frac{1}{2(2 \pi)^2 g_{YM2}^2} \int dV_{3} \left( (2 \pi)^2 \tilde{V}\left(\dot{\Phi}^{2}+\Phi^{2}\dot{\Omega}_{2}^{2}\right)+\frac{(g_{YM2}^2)^2}{4\tilde{V}}\left(\dot{\varphi} -(k+2N_{4})\cos\theta \dot{\phi}\right)^{2}\right)
\ee
Reading off the metric from the kinetic terms we find precisely the
Taub-Nut-like metric expected from field theory. The fact that the
moduli-space metric degenerates when $\tilde V$ becomes negative
is a strong indication that the geometry must be corrected
significantly inside the enhancon radius.

\item It is natural to contemplate generalization with more than 2 NS5
  branes and general linking numbers so that the flavors are charged
  more generally under the gauge group which has a product
  structure. Some preliminary discussion on this point is discussed in
  section 2.6.2 of \cite{Aharony:2009fc}. See also
  \cite{Witten:2009xu}. Analyzing these constructions in detail
  appears somewhat subtle, and will be left for future work.

\item It is also interesting to compare the IIA solution we reviewed
  here to the IIB solution discussed in
  \cite{Assel:2011xz,Assel:2012cj}. There are two main differences. 

One is that the solution of \cite{Assel:2011xz,Assel:2012cj} considers
only the gravity dual of the IR fixed point, whereas we are
considering the gravity dual of the full renormalization group flow
starting with gauge field theory in the ultraviolet. We will study the
intricacies of the renormalization group further in the following
sections. These issues are inaccessible in the solutions of
\cite{Assel:2011xz,Assel:2012cj}.

Another difference is the obvious one between the IIA and the IIB
solutions. These are related by T-duality along the Hopf fiber
direction of the ALE space. Usually, only one of the T-dual pair is
the preferred duality frame in the sense that the effective dynamics
is better encoded in the supergravity approximation.  Whether one
should or shouldn't T-dualize along the Hopf fiber of the $C^2/Z_{k'}$
orbifold has a lot to do with the size of $k'$. When $k'$ is large, it
makes good sense to T-dualize from IIA to IIB \cite{Mukhi:2002ck}.  In
the case where $k'=2$ as in the solution reviewed in this section, it
is more effective to work in the IIA frame. The full string theory
should, of course, encode all of the physics. 

\item The final step in constructing the solution is solving for the warp factor (\ref{harm}). Aside from the source term, (\ref{harm}) is linear. We can therefore break up $H(\vec y,\vec r)$ by writing
\be H(\vec y,\vec r) = 1 + H_1(\vec y,\vec r)+H_2(\vec y,\vec r) \ee
where
\beq 0 &=&  \left(\nabla^2_y+ \nabla^2_{TN} \right) H_1 + (2 \pi l_p)^6  Q_2 \delta^4(\vec y) \delta^4(\vec r) \label{harm1} \\
0 & = & 
 \left(\nabla^2_y+ \nabla^2_{TN} \right) H_2 + {l^2 V^4\over 2  r^4}  \delta^4(\vec y) \label{harm2}
\eeq
with the boundary condition that $H^1$ and $H^2$ decay at infinity.

$H_1(\vec y,\vec r)$ can be solved following \cite{Cherkis:2002ir}, with the only difference being some factor of $2$ arising from the $Z_2$ orbifold in $C^2/Z_2$ and $k$ from the Taub-NUT charge.  The solution is 
\beq H_1(y,r) &=& (2k) 32 \pi^2  l_p^6 Q_2 
\int dp \left({(py)^2 J_1(py) \over 4 \pi^2 y^3}\right)  H_{1p} \\
H_{1p} & = & c_p e^{-pr} {\cal U} \left(1 + {k p R_{11} \over 4}, 2, 2pr\right) \\
c_p & = & \left({\pi^2 \over 8} p^2 \Gamma\left( {p k R_{11} \over 4} \right) \right) \eeq

Using similar separation of variable technique, we have
  for $H_{2p}$,
\be H_2(r,y) = 2\int dp \left({(py)^2 J_1(py) \over 4 \pi^2 y^3}\right) H_{2p} \ee
with
\be \left(\partial_r^2 + {2 \over r} \partial_r  - {p^2 \over V}  \right) H_{2p}  = - {l^2 V^3 \over 2 r^4} \ee
which can formally be solved using the method of variation (See (1.5.7) of \cite{BenderOrszag}.)
\be H_{2p} = H_{2p}^{(2)}(r)\int_{r_1}^{r}  dr' \left(-{ l^2 V^3 \over 2 r^4} \right){H_{2p}^{(1)}(r')  \over W(r')} 
- H_{2p}^{(1)}(r)\int_{r_2}^{r}  dr' \left(-{ l^2 V^3 \over 2 r^4} \right) {H_{2p}^{(2)}(r')  \over W(r')} 
\ee
where
\be W(r') = H_{2p}^{(1)} (r') H'{}_{2p}^{(2)} (r') -H_{2p}^{(2)} (r') H'{}_{2p}^{(1)} (r') \ee
and
\beq H_{2p}^{(1)} & = & e^{-pr} {\cal U} \left(1+ {p k R_{11} \over 4},2, 2 p r\right) \\
H_{2p}^{(2)} & = & 
e^{-pr} {}_1 F_1 \left(1+ {p k R_{11} \over 4},2, 2 p r\right) 
\eeq
where $r_1$ and $r_2$ parameterizes the freedom to adjust the integration constant. In order to make the solution regular at $r=0$ and $r=\infty$, we set $r_1 = \infty$ and $r_2 = 0$. 

The Wronskian for these solutions can be written compactly as
\be W = H_{2p}^{(1)}(r) H_{2p}^{(2)}{}'(r) - H_{2p}^{(2)}(r) H_{2p}^{(1)}{}'(r)  = {2 \over  k p^2 R r^2 \Gamma\left({1 \over 4} k p R\right)} \ .  \ee

\item We can now scale out $\alpha'$ dependence by substituting
\beq 
r &=& \alpha' U \\
y & = & \alpha' Y \\
p & = & \alpha'^{-1} P
\eeq
as well as scaling
\be H_i(y,u) = \alpha'^{-2} h_i(Y,U)  \ . \label{scaling}\ee
Then, we find
\beq h_1(Y,U) &=& (2k) 32 \pi^2 g_{YM}^4   Q_2 
\int dP \left({(PY)^2 J_1(PY) \over 4 \pi^2 Y^3}\right)  h_{1P} \\
h_{1P} & = & C_P e^{-PU} {\cal U} \left(1 + {k P g_{YM}^2 \over 4}, 2, 2PU\right) \label{firsteq} \\
C_P & = &   {\pi^2 \over 8} P^2 \Gamma\left( {k P g_{YM}^2  \over 4} \right) \eeq
and
\be h_2(U,Y) =  2 (2 \pi)^4 g_{YM}^4  \left(N_4 + {k b_\infty \over 2}\right)^2\int dP \left({(PY)^2 J_1(PY) \over 4 \pi^2 Y^3}\right) h_{2p} \ee
\be h_{2p} =  h_{2P}^{(2)}(U)\int_{\infty}^{U}  dU' \left(-{  V^3 \over 2 U'^4} \right){h_{2P}^{(1)}(U')  \over w(U')}  \
- h_{2P}^{(1)}(U)\int_{0}^{U}  dU' \left(-{  V^3 \over 2 U'^4} \right) {h_{2P}^{(2)}(U')  \over w(U')} 
\ee
where
\be w(U') = h_{2P}^{(1)} (U') h'{}_{2P}^{(2)} (U') -h_{2P}^{(2)} (U') h'{}_{2P}^{(1)} (U') = {2 \over g_{YM}^2 k P^2 U'^2 \Gamma({1 \over 4} k P g_{YM}^2)} \ee
and
\beq h_{2P}^{(1)} & = & e^{-PU} {\cal U} \left(1+ {P k g_{YM}^2\over 4},2, 2 PU\right) \\
h_{2p}^{(2)} & = & 
e^{-PU} {}_1 F_1 \left(1+ {P k g_{YM}^2 \over 4},2, 2 P U \right)  \label{lasteq}
\eeq
The essential point to take away here is that the only place where $\alpha'$ appears is in (\ref{scaling}), and the decoupling $\alpha' \rightarrow 0$ has the effect of simply dropping the ``1'' in (\ref{scaling}) while keeping everything else in (\ref{firsteq})--(\ref{lasteq}) fixed. This will result in having the string frame metric having no dependence on $\alpha'$ aside from the overall normalization 
\be ds^2 = \alpha' ( \ldots )  \ee
as is conventional in gauge gravity correspondences. 

\item Large/small radius behavior of the warp factor

Now that we have worked out the warp factor in a reasonably explicit
form, we can explore their asymptotic behaviors. The large and small
radius behavior of $h_1(Y,U)$ is identical to what was found in
\cite{Cherkis:2002ir}. In particular, for large $U^2+Y^2$, we find
\be h_1(Y,U) \sim {12   \pi^2 g_{YM}^2  }{Q_2 \over (U^2+Y^2)^{5/2}} \ , \ee
which is the warp factor expect for D2-brane in $C^2/Z_2 \times R^3$. 

For $h_2(Y,U)$, we find
\be h_2(Y,U) \sim 12 \pi^2 g_{YM}^2   {{1 \over k} \left(N_4 + {k b_\infty \over 2}\right)^2   \over (U^2 + Y^2 )^{5/2}} \ . 
\ee
The numerator
\be Q_{bulk} = {1 \over k} \left(N_4 + {k b_\infty \over 2}\right)^2 \ee
can be interpreted as the bulk contribution to Maxwell charge so that
\be Q_{Maxwell} = Q_2 + Q_{bulk} = N_2 + b_\infty N_4 + {1 \over 4} b_\infty^2 k \ee

For small $U$ and $Y$, on the other hand, we find that
\be h_1(Y,U) = {64 \pi^2 k  g_{YM}^4 \over (Y^2 + 2k g_{YM}^2 U)^3} Q_2, \qquad
h_2(Y,U) = {64 \pi^2  \over g_{YM}^4 k^4} {\left(N_4 + {b k \over 2}\right)^2 \over Y^2}
\ee
which takes on somewhat more homogeneous form when we substitute
\be U = {Z^2 \over  2k g_{YM}^2} \ee
so that
\be h_1(Y,Z) = {64 \pi^2 k g_{YM}^4 \over (Y^2 + Z^2)^3} Q_2, \qquad
h_2(Y,Z) = {64 \pi^2 \over g_{YM}^4 k^4} {\left(N_4 + {b k \over 2}\right)^2 \over Y^2}
\ee

What we see is that $h_2$ sources a wall of charges localized at $Y=0$
which dominates when $Q_2=0$. If $Q_2$ is positive, however, $h_1$
dominates near $Y=Z=0$, and asympototes to an $AdS_4 \times S_7/Z_k$
geometry whose radius in Plank unit is $Q_2$ up to some finite
dimensionless factor.  If on the other hand $Q_2$ is negative but
$Q_2^{Maxwell}$ is positive, then the background will contain a
repulson singularity which one expects to be resolved by the standard
enhancon mechanism. What is interesting about this class of
background, however, is the fact that the enhancon mechanism can be
relevant even in the absence of repulson singularities, as we will
discuss further below.

\end{itemize}

\subsection{Holographic interpretation of the supergravity solution}

Now that we have worked out the supergravity solution in detail, let
us examine their basic properties. The D2 Maxwell charge was found to
be
\be Q_2^{Maxwell} = N_2 + b_\infty N_4 + {b_\infty^2 \over 4} k \ . \ee
Let us restrict our attention to the case where this charge is positive.

The D2 brane charge localized at the origin, on the other hand, was found to be
\be Q_2 = N_2 - {N_{4}^{2} \over k} \ . \ee
If $Q_2$ is positive, we find that the region near the origin
asymptotes to $AdS_4 \times S^7/Z_k$ with curvature of order $Q_2$, as was found in \cite{Aharony:2009fc,Bergman:2009zh}

If instead $Q_2$ takes a negative value while keeping $Q_2^{Maxwell}$
positive, we encounter a singularity of a repulson type.  If all the
objects giving rise to net negative $Q_2$ are allowed BPS objects
e.g. flux and discrete torsion, this repulson singularity is expected
to be resolved by the standard enhancon mechanism and ultimately give
rise to regular string dynamics \cite{Johnson:1999qt}. If $Q_2 \ll
-1$, the configuration like does not exist as a supersymmetric
state.\footnote{In some constructions like in \cite{Cottrell:2013asa},
  this may be related to dynamical breaking of supersymmetry, but such
  phenomena will not be the focus of this paper.} 

Let us examine the solution more closely for explicit choice of parameters. 

\begin{itemize}

\item As the first concrete example, let us set
\be b_\infty = {1 \over 2}, \qquad N_2 = 2m, \qquad N_4 = -m, \qquad k = N_6 = 3m \ . \ee
We will take $m$ to be some large but finite integer, so that the
supergravity solution is effective for a wide range of scales
\cite{Itzhaki:1998dd}. We then have
\be Q_2 = {5  \over 3} m> 0 , \qquad Q_2^{Maxwell} = {27  \over 16} m > 0 \ . \ee

What we propose to do now is to probe this geometry in the $\Phi \sim r$ coordinates using a D4 and an anti D4+D2 probes fixed at the origin in the $\vec y$ coordinates. This is a crude probe of the Coulomb branch of the corresponding field theory.

\begin{figure}
\centerline{\includegraphics[width=5in]{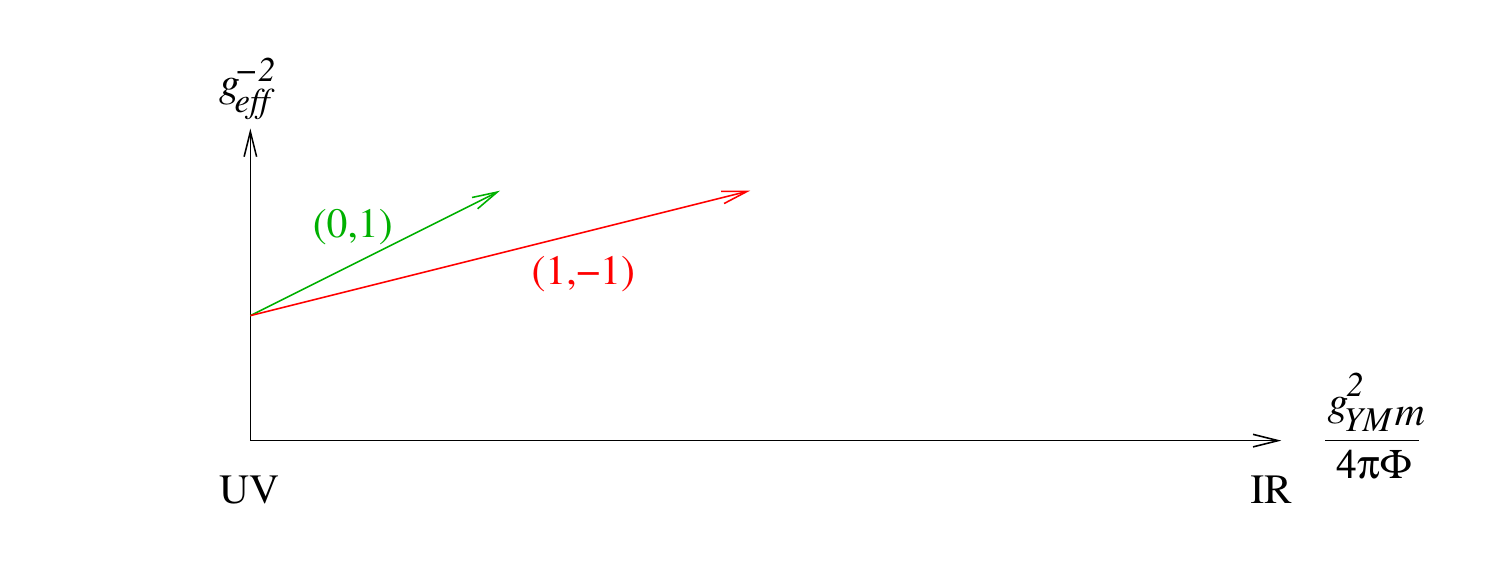}}
\caption{$g_{eff}^{-2}$ for D4 or anti D4 probe with some D2 charge, as a function of $\Phi$. The notation (D2,D4) represents the charge of the probe. For instance $(0,1)$ is a D4 probe, and $(1,-1)$ is an anti D4 probe with one unit of D2 charge. \label{figc}}
\end{figure}

If one plots $g_{eff1}^{-2}$ and $g_{eff2}^{-2}$ as given in
(\ref{geff1}) and (\ref{geff2}), it would look like what is
illustrated in figure \ref{figc}.  In particular, $g_{eff1}^{-2}$ and
$g_{eff2}^2$ remains positive. This is equivalent to the condition
that these probes satisfy (\ref{bound}) and remain BPS as they explore
the entire range of $\Phi$.

Nothing out of the ordinary happens, and the interpretation that this
gravity solution is describing the RG flow of ${\cal N}=4$ $U(2m)
\times U(m)$ system with $3m$ fundamentals in 2+1 dimensions, flowing
in the IR to a superconformal fixed point dual to an $AdS_4 \times
S^7/Z_{3m}$ geometry of radius $5m/3$ appears rather robust.

\item As a second example, consider setting

\be b_\infty = {1 \over 2}, \qquad N_2 = 2m, \qquad N_4 = m, \qquad k = N_6 = 3m\ .  \label{setting2} \ee

The only change is the sign of $N_4$.  It is easy to see that the this example is related to the previous one by the exchange of the position of the NS5-branes in the brane picture.

For this example, we have
\be Q_2 = {5 \over 3} m > 0, \qquad Q_2^{Maxwell} = {43 \over 16} m > 0 \ . \ee
So $Q_2$ is the same as in the previous example.

The $g_{eff1,2}^{-2}$ look very different in this case. In particular, the effective coupling diverges as one flows from large to small $\Phi$ for the D4 probe at 
\be  {g_{YM}^2 m \over 4 \pi \Phi_e} = {1 \over 4}  \ . 
\ee
This is illustrated in figure \ref{figd}. For $\Phi < \Phi_e$, the D4 probe is no longer BPS.

\begin{figure}
\centerline{\includegraphics[width=5in]{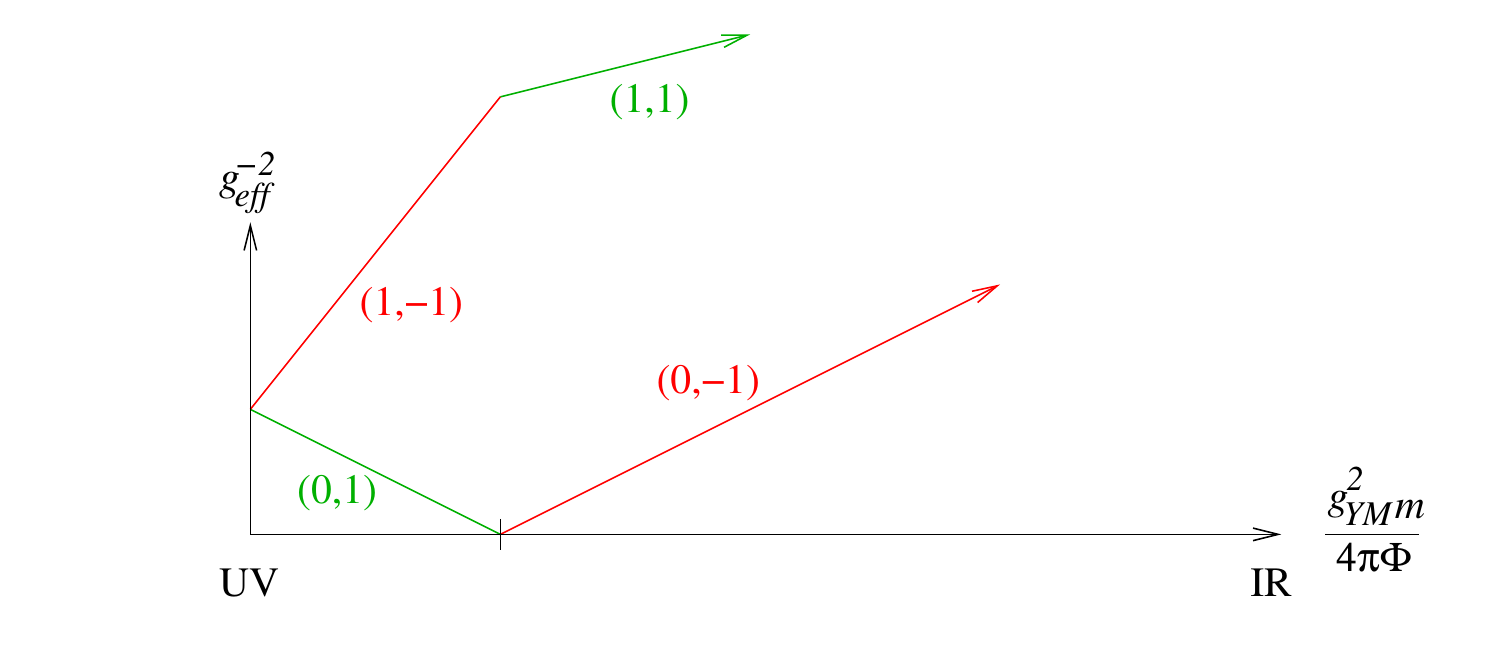}}
\caption{\label{figd} $g_{eff}^{-2}$ for probe branes in the second case (or with $N_4$ positive). Inside the enhancon radius different probes are BPS.}
\end{figure}

As it turns out, there are other probe which are BPS and can seemingly
probe the region $\Phi < \Phi_e$. In the region $\Phi< \Phi_e$, one
can use the $(1,1)$ and the $(0,-1)$ probes. These branes will probe the $AdS_4 \times S^7 / Z_{3m}$ geometry deep in the  small $\Phi$ region without any problems.

\item As the third example, let us consider the case
\be b_\infty = {1 \over 2}, \qquad N_2 = 7m, \qquad N_4 = -4m, \qquad k = N_6 = 3m \ee
so that 
\be Q_2 = {5 \over 3} m > 0, \qquad Q_2^{Maxwell} = {83\over 16} m > 0 \ . \ee
This time, we see the $(1,-1)$ probe cease to be BPS at 
\be  {g_{YM}^2 m \over 4 \pi \Phi_e} = {1 \over 10}  \ . 
\ee
We can continue to probe the region $\Phi < \Phi_e$ using probes with
charges $(-1,1)$ and $(2,-1)$. The $(2,-1)$ probe eventually ceases to
be BPS, but one can probe beyond that region using yet another set of
probes $(3,-1)$ and $(-2,1)$. This set of probes remain BPS and valid
all the way down to the origin in $\Phi$ space where the geometry
asymptotes to $AdS_4 \times S_7/Z_{3m}$.

\begin{figure}
\centerline{\includegraphics[width=5in]{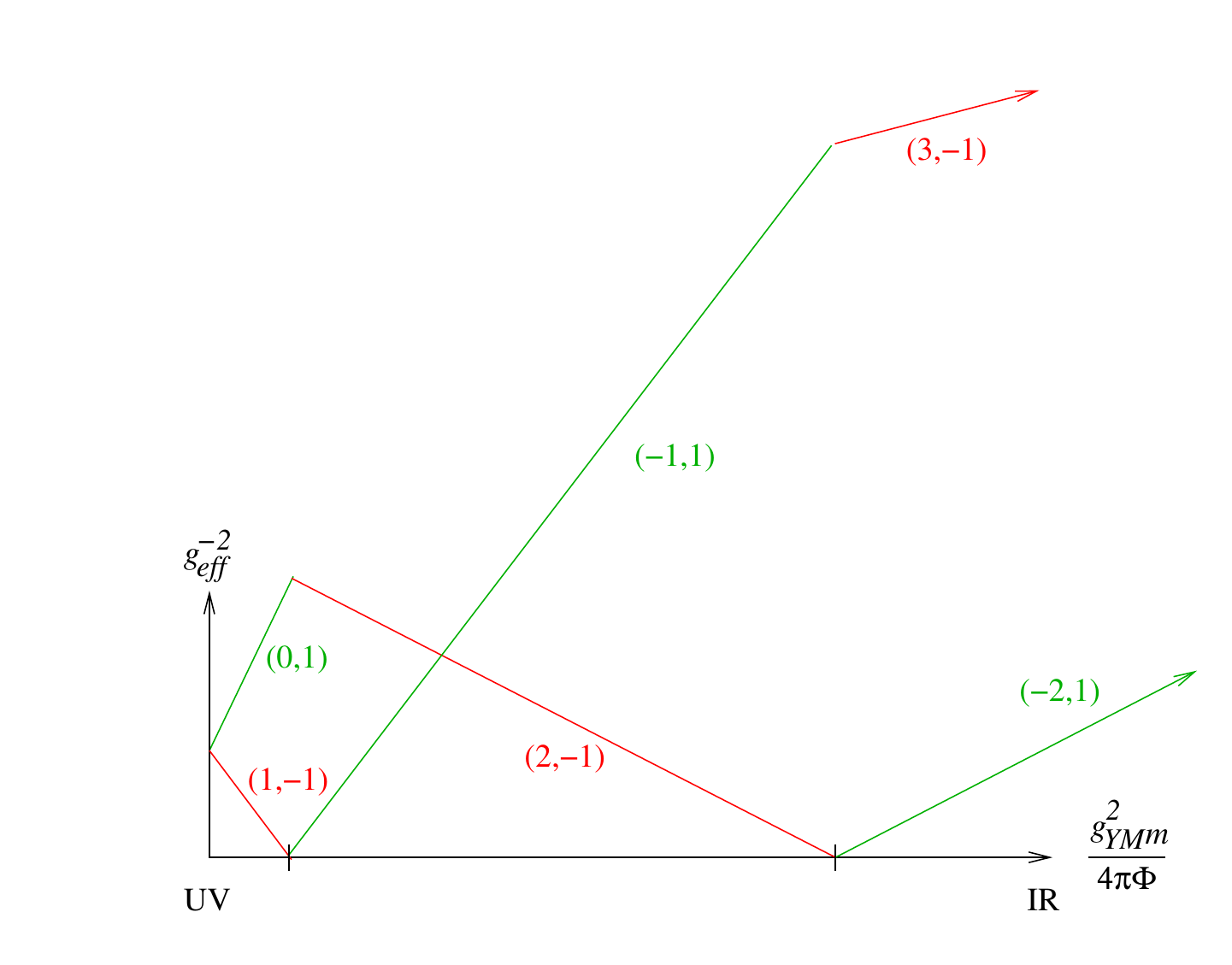}}
\caption{\label{fige} $g_{eff}^{-2}$ for probe branes in the third case. There are two enhancon radii and in each region there are different BPS probes.}
\end{figure}

\end{itemize}

What we seemingly have at our hand is a gravity solution dual to
$U(2m) \times U(m)$, $U(2m) \times U(3m)$, and $U(7m) \times U(3m)$
theories with 3 fundamentals charged under the $U(2m)$ or $U(7m)$, which
are free of repulson singularities, and seemingly all asymptoting to
an $AdS_4 \times S^7/Z_k$ geometry in the IR with radius $5m/3$. It is
tempting, as was suggested in \cite{Polchinski:2000mx} to regard these
backgrounds as exhibiting an analogue of duality cascade
\cite{Klebanov:2000hb}. This then amounts to claiming that the three
gauge theories listed above are related by Seiberg-like duality, and
all have the same superconformal field theory as the infra-red fixed
point.

There is however a main flaw in this argument. Seiberg's duality in
3+1 dimensions \cite{Seiberg:1994pq} and the related Aharony duality
in 2+1 dimensions \cite{Aharony:1997gp} are features of field theories
with 4 supercharges. For theories with 8 supercharges like the ones we
are considering, there is no established duality can be considered
analogous. In the absence of dualities, one can not conceive of a
duality cascade.

Another obvious difficulty in claiming that the three field theories considered as an example above are related by duality is the basic fact that the dimension and the structure of moduli space is completely incompatible. 

This issue was articulated explicitly in \cite{Aharony:2000pp} for
${\cal N}=2$ theories in 3+1 dimensions. In order to provide some
holographic interpretation to the supergravity solution like the one we
constructed in the last subsection, \cite{Aharony:2000pp} suggested
that the background is dual to some specific choice of vacuum on the
Coulomb branch where the effective rank of the field theory is
gradually reduced by Higgs mechanism.

This issue was further elaborated in \cite{Benini:2008ir} which
constructed the supergravity solutions that are interpretable as being
the dual of the ${\cal N}=2$ theory in 3+1 dimensions. By providing
explicit supergravity solution for generic vacuum on the Coulomb
branch, one can diagnose the hypothesis that the solutions found in
the previous subsection is interpretable as some specific choice among
the set of possible vacua. In this regard, the conclusion is somewhat
anti-climactic. As long as the scale of the vacuum expectation value
is greater than the scale set by $\Phi_e$, one can reliably interpret
the supergravity solution, but as one approach near the origin/root of
the Coulomb branch, the supergravity solution is suffering from being
unreliable on the account of tensionless brane objects nucleating at
the enhancon radius $\Phi_e$.

The crude diagnostic is that regions behind the first enhancon radius
appearing at scale $\Phi_e$ do not exist unless the theory is
sufficiently higgsed at scale exceeding $\Phi_e$. In the case of
${\cal N}=2$ theories in 3+1 dimensions, one can further argue that
$\Phi_e$ is the minimal allowed higgsing that is allowed due to
quantum corrections on the Coulomb branch which are analyzable using
the technology of Seiberg-Witten theory
\cite{Seiberg:1994rs,Seiberg:1994aj}.  In particular, one can identify
a special point on Coulomb branch called the ``baryonic root'' which
is a unique point where the Coulomb branch and the baryonic branch
meet \cite{Argyres:1996eh}. For the supergravity duals of ${\cal N}=2$
theories in 3+1 dimensions, \cite{Benini:2008ir} showed that the dual
of the baryonic root corresponds to arranging the fractional branes
exactly at $\Phi_e$ to screen the enhancon.

Our ultimate goal to study these issues for the case of 2+1
dimensions. There are few obstacles that one needs to overcome in
order to carry out this program in full. One is the fact that the
technology of Seiberg-Witten theory is not as developed in 2+1
dimensions. We need to map out the structure intersections of Coulomb
and Higgs branches for the ${\cal N}=4$ theories. There have been a
number of useful recent developments
e.g.\ \cite{Hashimoto:2014vpa,Hashimoto:2014nwa,Bullimore:2015lsa}
which we intend to exploit to develop this side of the story further.

In this article, we will take the first step in this program by
constructing supergravity solutions which screens the enhancon
singularity.  More specifically, we construct the analogues of the
explicitly higgsed solution of \cite{Benini:2008ir}. 

The essential conclusion we will arrive at is that gravity solution
which exhibits an enhancon, even in the absence of a repulson, should
be considered unreliable inside the enhancon radius. That certain
seemingly good supergravity solution is nonetheless unreliable because
of the behavior of probe branes may have far reaching impacts in
subjects such as black hole information paradox, since vacua with
fluxes and orbifold fixed points that give rise to these enhancon like
structures is rather ubiquitous in string theory. 

It is also useful to pause and note that the breakdown of supergravity
due to enhancon mechanism does not always need to happen. In fact, it
did not in the first example illustrated in figure \ref{figc}. The
condition for enhancons not to appear is for (\ref{geff1}) and
(\ref{geff2}) to both exhibit the IR free running. In other words,
\be k \ge  - 2 N_4 \ge 0 \ . \label{cond1} \ee
This is to be combined with the other requirement
\be Q_2 = N_2 - {N_4^2 \over k} > 0 \ . \label{cond2} \ee
These are the conditions that the supergravity solution is well behaved.

\begin{figure}
\centerline{\includegraphics{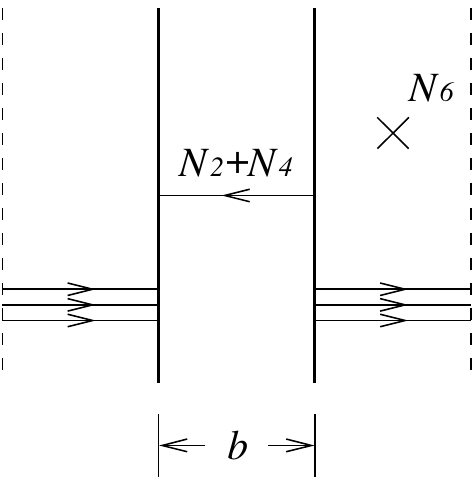}}
\caption{\label{fignonsusy} Hanany Witten type brane configuration for $N_2+ N_4< 0$.}
\end{figure}

Here, however, we encounter a curoius puzzle. In order for the brane configuration underlying the construction to preserve supersymmetry, we expect the condition
\be N_2 > - N_4 > 0 \label{cond3} \ee
to be satisfied.  If (\ref{cond3}) is violated, there will be some
anti D3 segements as is illustrated in figure \ref{fignonsusy} which one expects will lead to the complete breaking of
supersymmetry.

The conditions (\ref{cond1}) and (\ref{cond3}) combined is equivalent
to the criteria for the circular quiver to be of the ``good'' type (in
the classification of \cite{Gaiotto:2008ak}) as was specified by
conditions
\be \rho^T > \rho \ee
in (2.25) of \cite{Assel:2012cj} and $L \ge 0$. One can further show
that given (\ref{cond1}), condition (\ref{cond2}) is strictly weaker
than (\ref{cond3}). This however creates an interesting conundrum. The
supergravity solution satisfying (\ref{cond2}) but violating
(\ref{cond3}) appears to be perfectly sensible supersymmetric
background although we expect its field theory dual not to be
supersymmetric.  We will comment further on this puzzle in the Conclusions.

\section{Higgsing}

In this section, we construct a generalization of the supergravity
solution constructed in the previous section where we higgs the $U(N)
\times U(M)$ theory to $U(N) \times U(M-P) \times U(1)^{P}$ by moving
$P$ fractional branes away from the origin in the $\vec r$ direction
transverse to the D6-brane, which is taken to be positioned at the
origin.\footnote{This higgsing here refers to turning on the vacuuum
  expectation value for scalars in the vector multiplet and therefore
  corresponds to exploring the Coulomb branch, and should not be
  confused with exploring the Higgs branch.} Since these
configurations are BPS, the $P$ fractional branes an be positioned
arbitrarily in $\vec r$ space, but to keep the analysis simple, we
will only consider the case where the $P$ branes are distributed
uniformly along a spherical shell at some fixed radius $r_s$.

We will trace the construction of the unhiggsed solution by introducing the D6 brane first, followed by the D4 brane, followed by the D2 brane.

Let us therefore start with the $R^{1,2} \times (C^2/Z_2) \times TN_k$ geometry in 11 dimensions, reduced to type IIA on the Hopf fiber of $TN_k$.

When $P$ D4-branes move in the direction transverse to the D6-branes,
the form of the self-dual 4-form is expected to be modified to account
for the D4 sources. Since the D4-brane is wrapping the collapsed
2-cycle of the ALE, we expect $G_4$ to maintain the form of being
$\omega_2$ wedged with some 2-form on $TN_k$. Locally, away from the
position of the D4 sources, $G_4$ in $(C^2/Z_k) \times TN_k$ should be
self dual.

A D4-brane at a generic point transverse to the D6-brane is expected
to carry a single unit of D4 brane charge, a single unit of D4 Page
charge, and $b(r)$ unit of D2 brane charge.

Since the distribution of D4-branes are spherical and codimension one along the orbifold fixed point, we expect the M-theory four form sourced by them to have the same general form as what we discussed in previous section with jump in $l$ and $\alpha$ at the spherical shell. The jump in $l$ and $\alpha$ should account for the brane and Page charges locally supported on the shell. In other words, we expect
\be G_4 = l(r) \omega_2 \wedge d (V \sigma_3) \ee
but with 
\be l =
\left\{ \begin{array}{ll} 
 - (2 \pi)^2  g_s l_s^3 \left({1 \over 2} k b_\infty +  N_4 \right) & (r > r_s) \\
- (2 \pi)^2  g_s l_s^3 \left({1 \over 2} k b_\infty +  N_4- { P \over V(r_s)} \right)  & (r < r_s)
\end{array} \right. \ee
with
\be N_4  =  M - N  \ . \ee
This discontinuity in $G_4$ accounts for the brane source.

We also need to know the $r$ dependence of $B_2$. This is constrained by the Page charge localized on the shell.  We find
\be \alpha  =  \left\{ \begin{array}{ll}
 (2 \pi)^2  g_s l_s^3 N_4 & (r > r_s) \\
(2 \pi)^2  g_s l_s^3 (N_4-P) & (r < r_s) \end{array}\right. \ . \ee
so that

\be b(r) = 
\left\{ \begin{array}{ll} b_\infty V(r) - {2 N_4 \over k} (1 - V(r))  & (r > r_s) \\
 b_\infty V(r) - {2 N_4 \over k} (1 - V(r)) + {2 P \over k} \left(1 - {V \over V(r_s)}\right) & (r < r_s) \end{array}\right. \ . \label{bofr}\ee
Note that $b(r)$ is continuous at $r=r_s$.  Also, 
\be b_0 = -{2 (N_4 - P) \over k} \ . \ee

What remains then is to compute the warp factor by solving
(\ref{harm1}) and (\ref{harm2}) suitably generalized to account for
the jump in 4-form flux as well as the D2-brane charge $b(r_s) P$
induced by $B_2$ field threading the D4-brane. The ``Maxwell charge at
radius $r$ is
\be Q_2^{Maxwell}(r) = \left\{
\begin{array}{ll} 
N_2 + b(r) (N_4-P) + {b(r)^2 k \over 4} & (r < r_s)  \\
N_2 + b^>(r) N_4 + {b(r)^2 k \over 4} & (r > r_s) \end{array}\right.
\ee
for $b(r)$ given in (\ref{bofr}) 
so that 
\be Q_2 = Q_2^{Maxwell}(0) = N_2 - {(N_4-P)^2 \over k} \ . \ee
We therefore see that effectively, at $r=r_s$, the supergravity
solution transitions from $U(N)\times U(M)$ theory to $U(N) \times
U(M-P)$ theory, as one would expect from the  standard Higgs mechanism.

Let us examine how this affects the RG flow in the specific example
considered in figure \ref{figd}. Below $\Phi < \Phi_s = r_s/\alpha'$,
the running changes to
\beq
{1 \over  g_{eff1}^2(\Phi)} &=& b_\infty  - { g_{YM}^2 (N_4 -P)\over 2 \pi \Phi} \ , \label{geff1a} \\
{1 \over  g_{eff2}^2(\Phi)} &=& (1 - b_\infty) + {g_{YM}^2(k + 2 N_4-2P) \over 4 \pi \Phi} \label{geff2a} \ .
\eeq

\begin{figure}
\centerline{\includegraphics[width=5in]{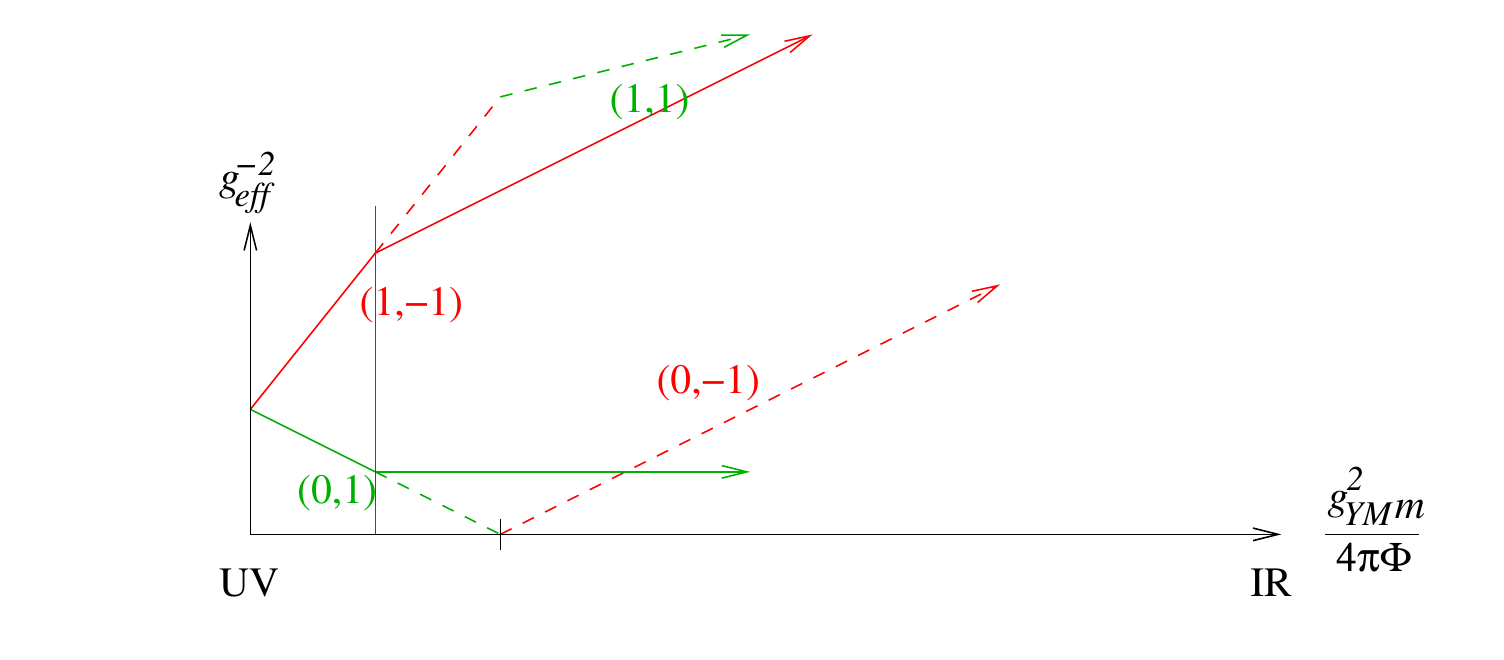}}
\caption{\label{figf} $g_{eff}^{-2}$ from figure 4  modified by enhancon screening  with $P=N_4$.}
\centerline{\includegraphics[width=5in]{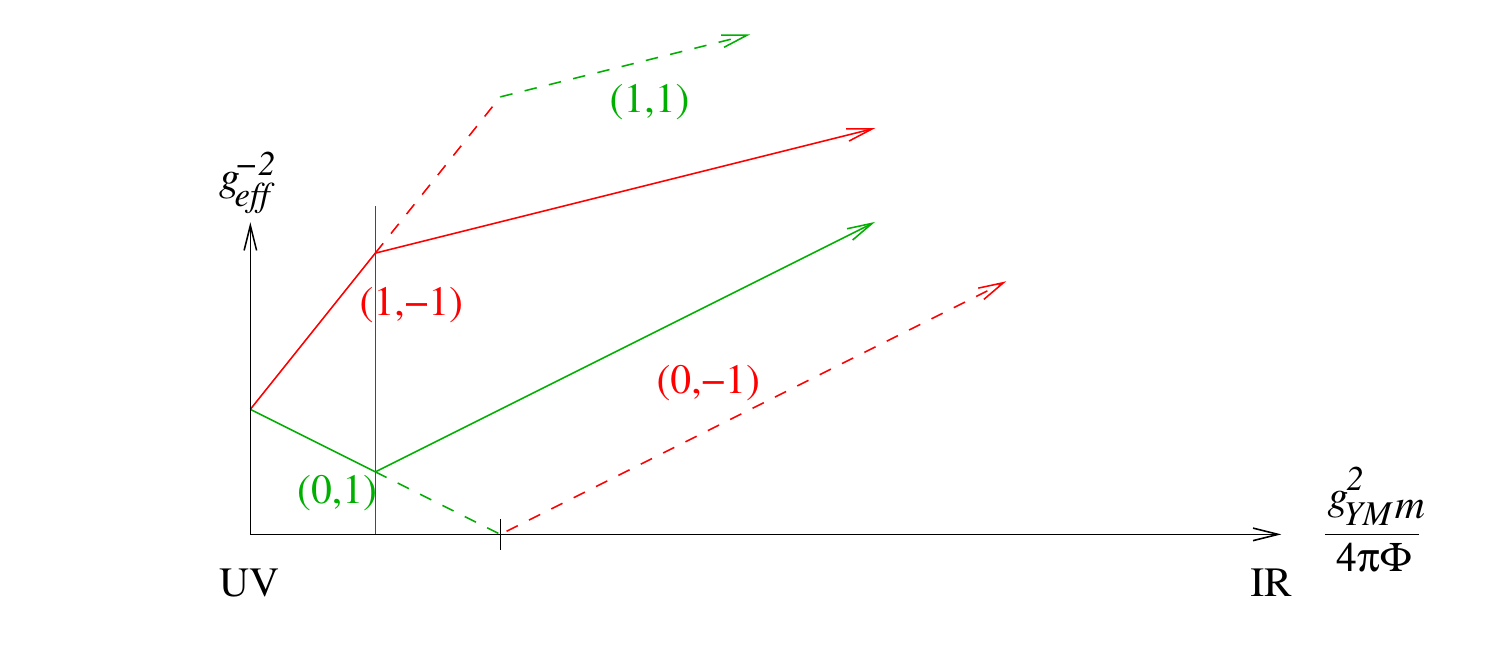}}
\caption{\label{figg}  $g_{eff}^{-2}$ from figure 4 modified by enhancon screening with $P=2N_4$.}
\end{figure}

For $P = N_4$, the modified running of the coupling looks like what is
illustrated in figure \ref{figf}.  We can also take $P = 2 N_4$
(assuming $2 N_4 \le k$), for which the running of coupling looks like
figure \ref{figg}. This latter choice mimics the beta function
coefficient of the naive cascade dual. Clearly, there are other
possible choice of $P$ that will eliminate the enhancon. Any $P$ in the range
\be N_4 \le P \le N_4  + {k \over 2} \ee
and $\Phi_s > \Phi_e$ would do. These are by no means intended to be
an exhaustive set of enhancon screening configurations. One can easily
envision relaxing the ansatz that the D4's arrange themselves in a
spherically symmetric fashion. The point of this exercise is to show
that 1) an explicit supergravity solution corresponding to specific
points on Coulomb branch like the one we considered is possible in
practice, and that 2) there are multitude of enhancon shielding
configurations.

Of course, as $\Phi_s$ approaches and crosses $\Phi_e$, the enhancon
returns. The most sensible and conservative interpretation is that any
feature encoded in supergravity in the region $\Phi < \Phi_e$ when the
enhancon is unshielded is unreliable.

It would be gratifying, on the other hand, if the appearance of
enhancons when $\Phi_s$ approaches $\Phi_e$ is signaling that the
geometry of the Coulomb branch is modified such that one simply can
not un-higgs beyond $\Phi_s = \Phi_e$. A picture roughly along these
lines was suggested by \cite{Benini:2008ir} relying mostly on the
structure of quantum exact Coulomb branch geometry inferred using the
Seiberg-Witten technique.  It would be interesting to attempt to close
this gap by studying the quantum corrected Coulomb branch geometry for
theories in 2+1 dimensions. This issue is currently under
investigation and will be reported in a separate publication. 

Another interesting question is whether there is a specific
supergravity solution (perhaps among the class considered in this
article, or its suitable generalization) which corresponds to special
points on the Coulomb branch such as the baryonic root
\cite{Argyres:1996eh}. The baryonic root is a point on moduli space
and is, in a manner of speaking, the natural one to identify as the
``origin'' of the Coulomb branch. It is the point where the baryonic
branch and the Coulomb branch meet. The branch and geometric structure
of the full moduli space and its supergravity manifestation is
currently under investigation, which we hope to more thoroughly address in
future work.

\section{The case of $k=0$}

In this section, we will extend the construction and the analysis of
the supergravity dual of $U(N) \times U(M)$ theory but with no flavor.
Some of these issues was discussed briefly in sections 2.3 and 2.4 of
\cite{Aharony:2009fc} but we will elaborate further on some of the
subtleties which were not highlighted there.  The $(C^2/Z_2) \times
TN_k$ geometry at the root of the construction is now modified to
$(C^2/Z_2) \times R^3 \times S^1$. The fact that fibration over $S^1$
is trivial simplifies issues such as disambiguation of Page,
brane, and Maxwell charges. This also implies that the details of
charge quantization will be different from what we saw in the previous
sections, which one might have anticipated from appearance of various
factors of $k^{-1}$.  We will encounter various singularities which we
will examine in some detail.

When $N=M$, the supergravity solution is simply that of $AdS_4 \times
S^7/Z_2$. Let us begin by formulating an ansatz for the supergravity
solution with some fractional branes present so that $N \ne M$.

We begin with an ansatz in M-theory where we warp the $R^{1,2} \times (C^2/Z_2) \times R^3 \times S^1$ geometry with an ansatz of the form
\be ds^2 = H^{-2/3} (-dt^2 + dx_1^2 + dx_2^2) + H^{1/3}(ds^2_{C^2/Z_2} + ds_{R^3 \times S_1}^2) \ee
Unlike the $k\ne 0$ case, we set 
\be x^{11} = R_{11} \psi \ee
and let $\psi$ have periodicity $2 \pi$. We take the ansatz for the M-theory 3-form to be
\be C_3 = H^{-1} dt \wedge dx_1 \wedge dx_2 + C_3^{SD} \ , \ee
with
\be C_3^{SD} = l \omega_2 \wedge \left({R_{11} \over 2r} d \psi + \cos \theta d \phi \right) + \alpha \omega_2 \wedge d \psi \ee
The four form field strength
\be G_4^{SD} = -l \omega_2 \wedge \left({R_{11} \over 2r^2} dr \wedge  d \psi + \sin \theta d \theta \wedge d \phi \right) 
\ee
is self-dual on $(C_2/Z_2) \times R^3 \times S^1$, but is not normalizable.  This is the first indication that something subtle is happening.  In fact, the equation for the warp factor is the $k \rightarrow 0$ limit of (\ref{harm}) and reads
\be 0 =  \left(\nabla^2_y+ \nabla^2_{TN} \right) H + {l^2 \over 2  r^4}  \delta^4(\vec y)
+ (2 \pi l_p)^6  Q_2 \delta^4(\vec y) \delta^4(\vec r)
\label{k0harm} \ , \ee
from which one can immediately infer that the bulk charge
\be \int G_4^{SD} \wedge G_4^{SD}  \sim  R_{11} \int d^3 r\, {l^2 \over 2 r^4} \label{divergebulk} \ee
diverges near $r=0$.  This divergence is addressed by having the
enhancon mechanism excise the region near $r=0$, which was how this
solution was presented in figure 2 of \cite{Aharony:2009fc}, but it
would be nice to see that in a more controlled manner. This is what we
will work out in this section. We will in fact see that most of these
features are hidden far inside the enhancon radius and is therefore,
in many ways, moot.

The reduction to IIA of our ansatz takes the form
\beq
ds^2 & = & H^{-1/2} (-dt^2 + dx_1^2 +dx_2^2) + H^{1/2} (ds_{ALE}^2 + ds_{R^3}^2) \\
A_1 & = & 0 \\
A_3 & = & -(H^{-1}-1) dt \wedge dx_1 \wedge dx_2 - l \omega_2 \wedge \cos \theta d \phi,\\
B_2 & = & -(2 \pi l_s)^2 b \omega_2, \qquad b = {1 \over (2 \pi l_s)^2} \left({l \over r} + {\alpha \over g_s l_s}\right), \\
e^{\phi} & = & g_s H^{1/4} 
\eeq

Quantization of D4 charges is rather straight forward. In particular, we find
\be 2 \pi l = (2 \pi l_s)^3 g_s N_4 , \qquad 2 \pi  \alpha = g_s (2 \pi l_s)^3 b_\infty \ . \ee
In terms of $N_4$,  $b_\infty$, $g_{YM}^2=g_s l_s^{-1}$, and $2 \pi \Phi = \alpha'^{-1} r$ we can write
\be b(r) = b_\infty + {g_{YM}^2 N_4 \over 2 \pi \Phi} \ee
which is manifestly dimensionless. 

The quantization of D2 charge and its relation to $Q_2$ is subtle
because of the naively divergent bulk charge (\ref{divergebulk}). We
can, however, read off the analogue of (\ref{geff}) by analyzing the
leading derivative term in the expansion  of the D4 and anti D4 brane probe DBI action.
\be
{1 \over  g_{eff1,2}^2(\Phi)} = (n \pm b_\infty) \mp {g_{YM}^2 ( 2 N_4) \over 4 \pi \Phi} \
, \ee
which naively leads to the kind of ``cascade'' illustrated in figure 1.a of \cite{Aharony:2009fc}, except that, as stressed throughout this article, there are no cascades for ${\cal N}=4$ theories in 2+1 dimensions.

To address the issue of relating $Q_2$ to quantized charges, it is
useful to consider the case of $U(N) \times U(M=N+P)$ at the point on
Coulomb branch where gauge group is broken to $U(N) \times U(N) \times
U(1)^{P}$ by $P$ fractional branes forming an (approximately)
spherical shell at a radius $r_s$ in $R^3$. In such a setup, one
expects the solution for $r < r_s$ to precisely be $AdS_4 \times
S^7/Z_2$ in M-theory, dimensionally reduced to $AdS_4 \times CP_3 /
Z_2$ in IIA with $N=N_2$ units of D2 charge. We also expect $l=0$ in the region $r < r_s$. At $r=r_s$, there are $P$ D4's each carrying one unit of D4 charge and $b(r_s)$ units of D2 charge. 

In order for $b(r_s)$ to be a meaningful concept, we need $b(r)$ to be continuous at $r=r_s$. That was found to be the case when $k \ne 0$ by requiring the Page and the brane charge to jump by an appropriate amount in the previous section. Here, we do not have the same independent constraint on continuity of $b(r)$, but let us impose that as necessary condition to make the D2 brane charge carried by the $P$ D4-branes well defined. 

This then implies that
\be l = \left\{ \begin{array}{ll} 0 & (r < r_s) \\ {1 \over 2 \pi} (2 \pi l_s)^3 g_s P & (r > r_s) \end{array} \right. \ee
and
\be b(r) = \left\{ \begin{array}{ll} b_\infty - {g_s l_s P \over r_s} & (r < r_s) \\ 
b_\infty - {g_s l_s P \over r} & (r > r_s)\end{array} \right. \ . \ee

\begin{figure}
\centerline{\includegraphics[width=3in]{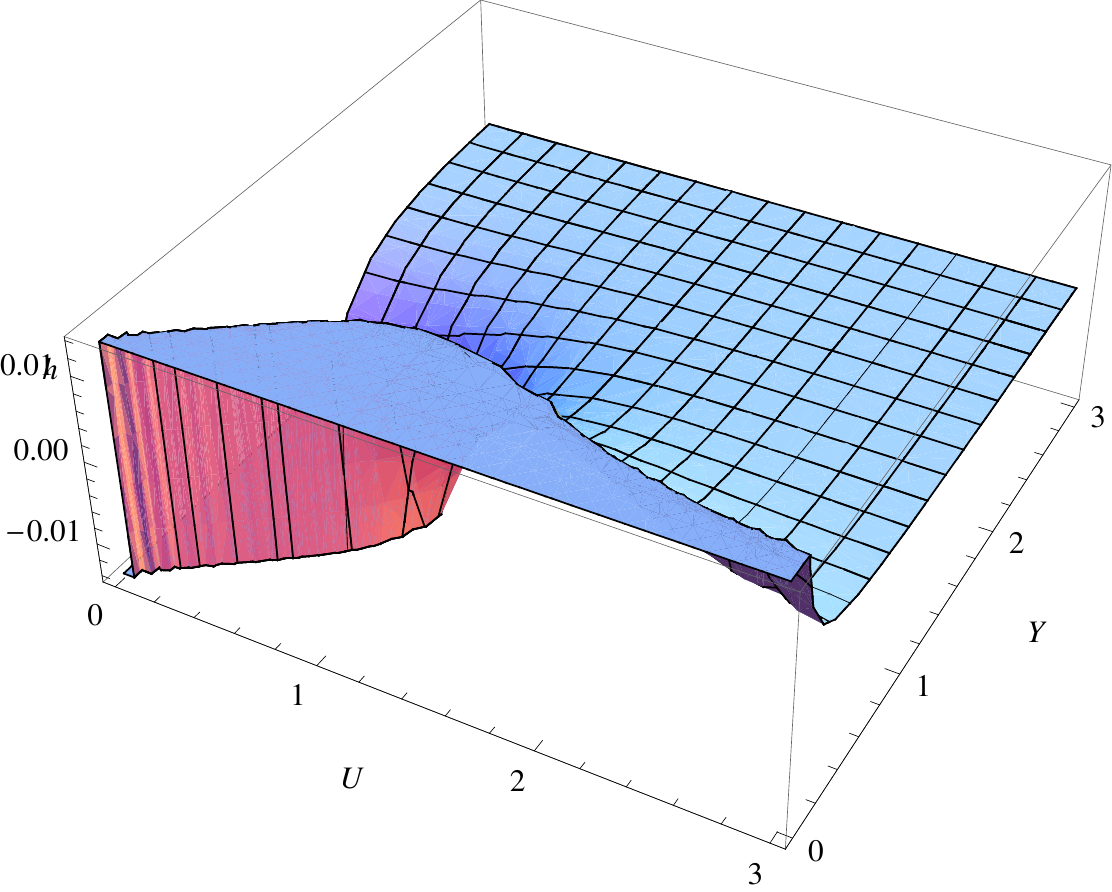}}
\caption{\label{figwarp} The solution to the harmonic equation in the decoupling
limit, for $k=0$. This figure originally appeared as figure 2 of \cite{Aharony:2009fc}}
\end{figure}

This spherical shell of D4 at radius $r_s$ regulates the bulk charge to
\be 2 \pi R_{11} \int_{r > r_s}  d^3 r {l^2 \over 2 r^4} = {(2 \pi)^2  l^2 R \over r} = (2 \pi l_s)^6 g_s^2 \left({P^2 g_s l_s   \over r}\right)\ee
Note that the sum of D2  brane and bulk charges,
\be Q_2^{Maxwell} = N_2 + b(r_s) P + {P^2 R  \over r_s}
= N_2 + b_\infty P  \ee
is happily independent of $r_s$.  On the other hand, both the brane
charge and the bulk charge diverge as $r_s$ approaches
zero. Introducing the spherical shell of radius $r_s$ is therefore an
effective way to regularize this divergence. One can in principle compute the warp factor and consider taking $r_s$ to zero. This will then give the same warp factor that was computed in figure 2 of \cite{Aharony:2009fc}  which we reproduce in figure \ref{figwarp}.

There is however one critical issue which requires attention. As one
 sends $r_s$ to zero, the flux of D2 charge at $r=r_s$
\be N_2 + b(r_s) P = N_2 + b_\infty P - {g_s l_s P \over r_s} \ee
turns negative at
\be r_s^* = {g_s l_s P \over N_2+ b P} = \alpha' {g_{YM}^2 P \over N_2 + b_\infty P} \ .  \ee
This is a repulson singularity, and signals that there when $r_s < r_s^*$, one shouldn't trust the supergravity solution to be capturing the physics of the field theory dual.  Note that the scaling with respect to $\alpha'$ is such that 
\be 2 \pi \Phi_s^* = {1 \over \alpha'} r_s^* = {g_{YM}^2 P \over N_2 + b_\infty P}   \ee
is finite in the $\alpha' \rightarrow 0$ limit.

\begin{figure}
\centerline{\includegraphics[width=5in]{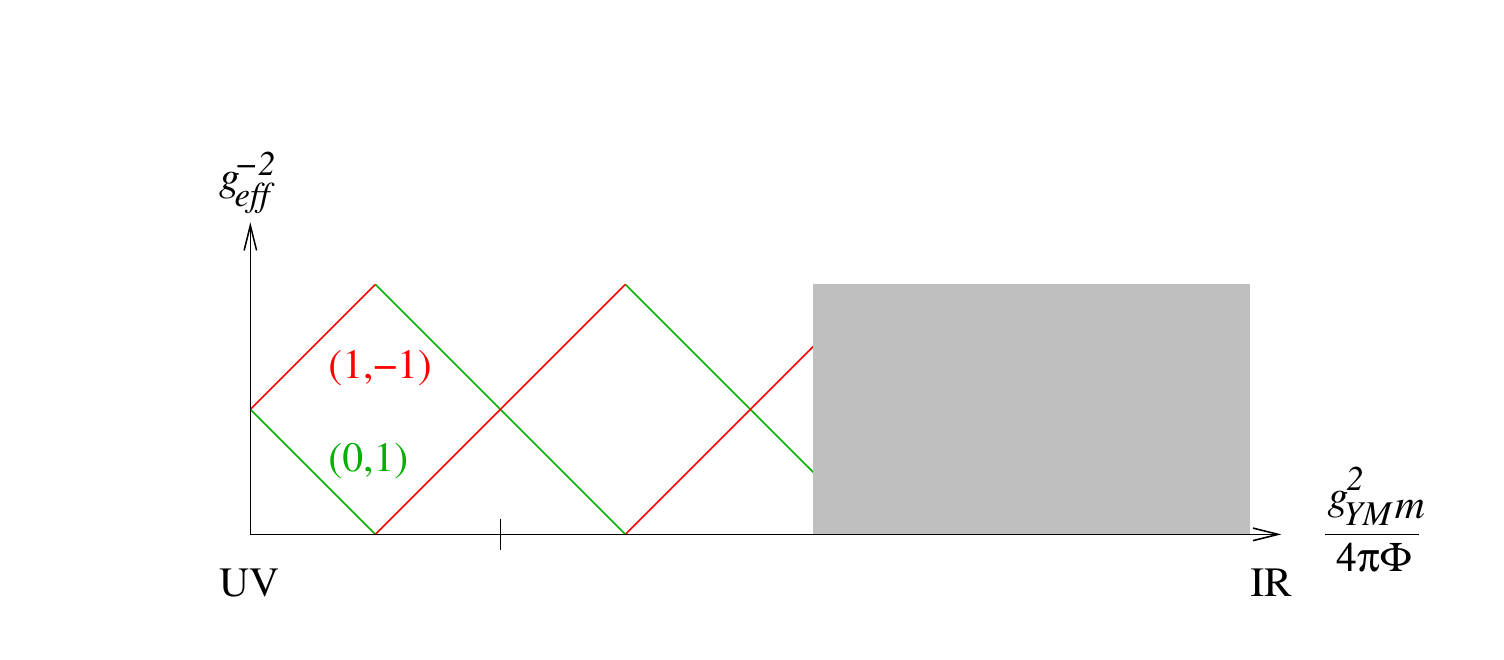}}
\caption{\label{figh} Cascade-like diagram. Repulson singularity is shaded in gray.}
\centerline{\includegraphics[width=5in]{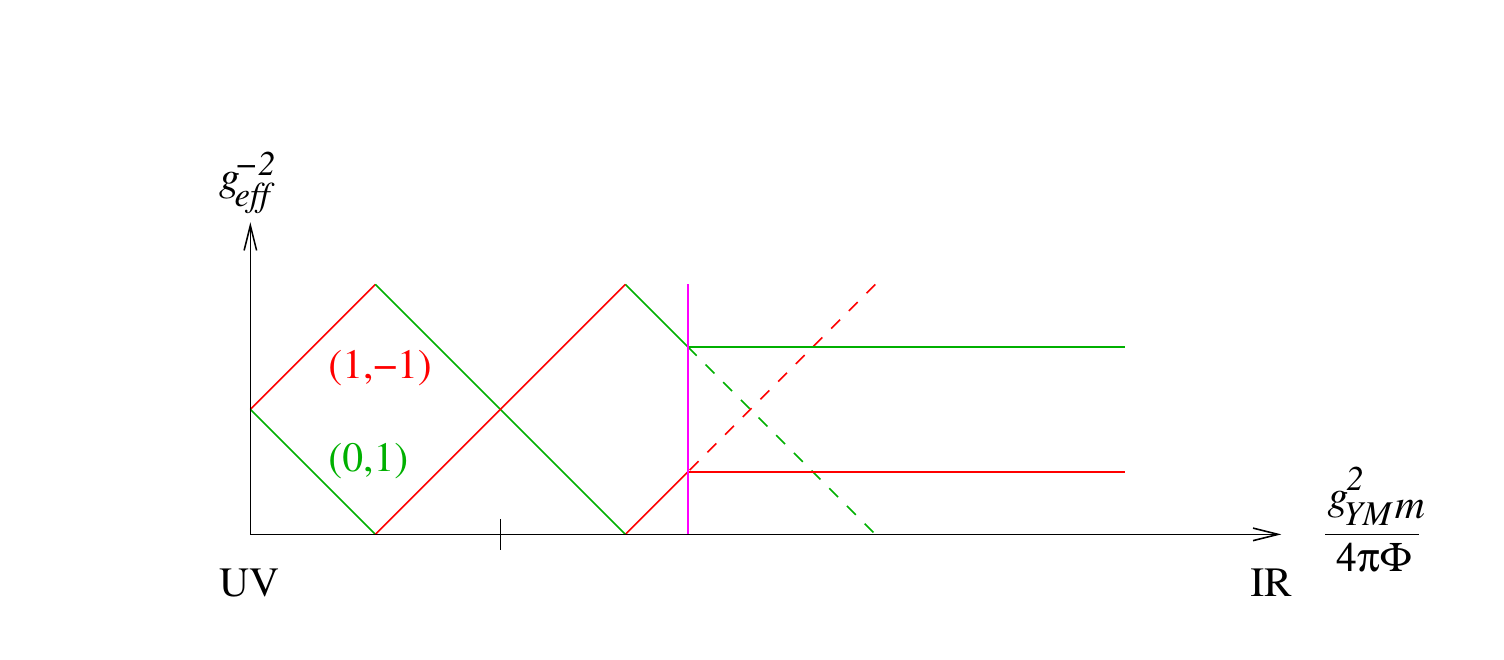}}
\caption{\label{figi} Cascade-like diagram with shielded repulson singularity.}
\centerline{\includegraphics[width=5in]{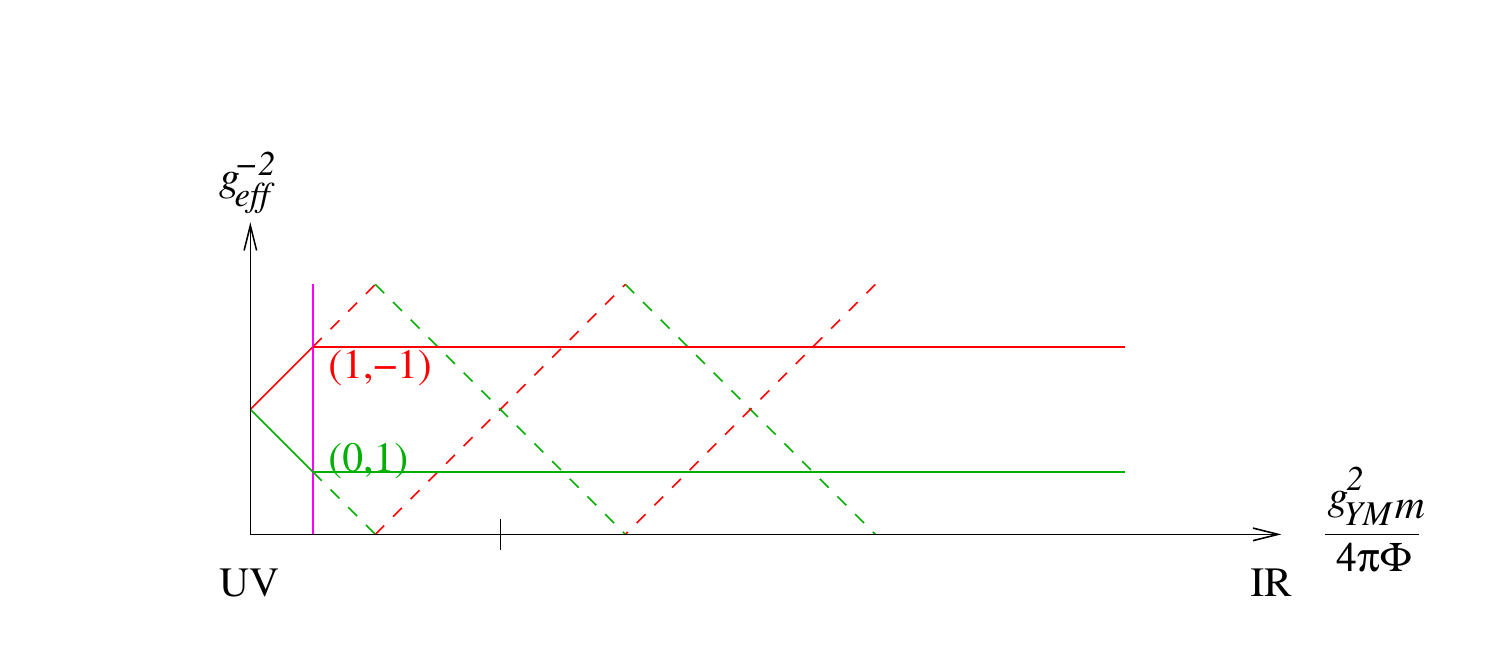}}
\caption{\label{figj} Trustworthy solution with all enhancon radii shielded.}
\end{figure}

One can visualize the renormalization group by drawing the cascade-like diagram, for instance, for the case of $N_2=7m$ and $N_4=4m$ as in figure \ref{figh}. The region
\be {g_{YM}^2 m \over 4 \pi \Phi} >  {g_{YM}^2 m \over 4 \pi \Phi_s^*} = {9 \over 8} \ee
behind the repulson singularity is shaded in grey.

One can shield the repulson singularity by smearing the $P=N_4=4m$ D4-branes in a spherical shell with radius
\be \Phi_s > \Phi_s^* \ee
as is illustrated in figure \ref{figi}. It might be tempting to conclude that as long as $\Phi_s > \Phi_s^*$, the supergravity solution is reliably capturing the dynamics of the gauge theory in the gravity dual description. 

It should be noted, however, that the outer-most enhancon, where one of $g_{eff}^{-1}(\Phi)$ vanishes, occurs at 
\be   { g_{YM}^2 N_4 \over 4 \pi\Phi_e} = {b_\infty \over 2} = {1 \over 4} \ . \ee
For the same reason as was discussed in the previous section,
supergravity solution in the region inside the outer-most enhancon
radius should not be considered reliable. In order to obtain a
reliable spherically symmetric supergravity solution, we must set the
radius $\Phi_s$ of the shell to be greater than the enhancon radius
$\Phi_e$ as is illustrated in figure \ref{figj}. It may turn out that
it simply meaningless to set $\Phi_s$ to be smaller than $\Phi_e$. It
would be interesting to find corroboration to this possibility from the
field theory side. At the level of supergravity solution, however,
upon setting $\Phi_s < \Phi_e$, one should treat the region $\Phi_s <
\Phi < \Phi_e$ as unreliable.

One can show that the repulson radius $\Phi_s^*$ is always smaller than the enhancon radius $\Phi_e$. This follows from the inequality
\be 2 \pi \Phi_e = {g_{YM}^2 P \over b_\infty} > {g_{YM}^2 P \over N_2 + b_\infty P} = 2 \pi \Phi_s^* \ee
or equivalently
\be N_2 + b_\infty P > b_\infty \ee
which can easily be seen to be always true for $N_2>0$, $P>0$, and $0 \le b_\infty \le 1$. 

\section{Conclusion}

In this article, we reviewed the construction \cite{Aharony:2009fc}
of supergravity duals of ${\cal N}=4$ field theories in 2+1 dimensions
with gauge group $U(N) \times U(M)$ arising from taking the decoupling
limit of brane construction illustrated in figure \ref{figa}. We then
scrutinized the regime of validity of the supergravity solution, and
highlighted the fact that
\begin{enumerate}
\item an enhancon can appear even in the absence of repulson singularities, such as in the examples illustrated in figures \ref{figd} and \ref{fige}, and 

\item at the enhancon radius, supergravity as an effective field theory breaks down because of the existence of tensionless brane objects. As such, the supergravity solution in the region inside the enhancon radius should be considered unreliable. 

\end{enumerate}

This implies then that aside from the somewhat restricted class of models
satisfying the constraints (\ref{cond1}) and (\ref{cond2}), only a
small region of the supergravity solution is reliable. In the region
where the supergravity ceases to be an effective low energy theory,
one expects qualitatively different dynamics than that which is
naively implied by the gravity solution. The general expectation is
that string and quantum corrections plays an important role. It would
be very interesting to verify this expectation on the field theory
side, to develop some sense on when and how gravity as an effective
theory breaks down in string theory. 

We also constructed a generalization of \cite{Aharony:2009fc}
corresponding to specific points on Coulomb branch where the
fractional branes are configured in an approximately spherically
symmetric distribution (which is reliable in the limit that the number
of fractional branes is large.) When the shell of fractional branes is
larger than the enhancon radius, all the singularities are screened
and the supergravity solution is globally reliable. This then suggests
that dynamically interesting things happen as the radius of the shell
approaches the enhancon radius. Unfortunately, it is not possible to
extract what that dynamics is from gravity alone, but perhaps some
information can be extracted from careful consideration of full string
theory on one side, and a detailed analysis on the field theory side.

For technical reasons, we restricted our attention to spherically
symmetric and smooth distribution of the fractional branes, but
solutions corresponding to arbitrary, discrete distribution of the
fractional branes should also exist. That is simply an exercise in
supergravity. We hope to address this point in the near future.  With
sufficient higgsing, one expects the enhancons to be shielded, giving
rise to a reliable supergravity dual.

A different way to regulate the dynamics of ``bad'' theories by
twisting the geometry to modify the scaling dimension of unitarity
violating magnetic monopole operators was suggsted in
\cite{Yaakov:2013fza}.\footnote{We thank Itamar Yaakov for very useful
  discussions on this point.} It would be interesting to see if a
gravity interpretation to the modifications invoked in
\cite{Yaakov:2013fza} can be identified, but we leave that question
for future work.

Finally, we took a closer look at the case with no flavors.  The case
without flavor has subtle differences to the flavored case with
regards to the details of charge quantization. The solution was found
to contain a singularity of a repulson type. This singularity can be
screened by higgsing along similar lines as what was done for the
flavored case. Nonetheless, the repulson singularity is always
surrounded by an enhancon singularity, and since one does not expect
supergravity features inside the enhancon radius on general grounds as
discussed repeatedly above, we do not expect to attribute much physics
to the repulson. On the other hand, one does expect interesting
physics, both on field theory and on gravity side, at the outer most
enhancon radius. 

The issue at the heart of this discussion is the condition for and
extent to which the geometry of the region of space-time inside the
enhancon is physically meaningful. Closely related issue was
discussed, for instance, in \cite{Polchinski:2000mx,Johnson:2001us}
where it is argued that as long as some probe can penetrate the region
inside the enhancon, the geometry must be reliable. Taking this
statement literally, however, would lead to the conclusion that a
``bad'' theory flows under renormalization group flow to a ``good''
``Seiberg-dual'' which is incompatible with the counting of moduli
space. A pragmatic point of view to take for the time being is to
interpret the enhancons as a hint that non-trivial dynamics could
dramatically correct the classical supergravity expectations, and
subject the system to more careful test from both field theory and
string theory sides to settle the issue. 

One surprising result we find is the mismatch in strength
  between regularity condition (\ref{cond2}) on supergravity sidde and
  (\ref{cond3}) based on expectation from brane construction.  As
  (\ref{cond2}) is weaker than (\ref{cond3}), one can satisfy the
  former while violating the latter. We believe this is a result of
  supergravity failing to account for higher curvature or quantum
  effects despite the fact that there are no obvious indication that
  supergravity, as an effective theory, is breaking down. This issue
  clearly deserves further consideration.

It would also be interesting to explore these solutions further, and
attempt to construct, to the extent that it is possible, solutions
corresponding to these theories at specific but generic points on
Higgs and Coulomb branches. It would also be instructive to carefully
examine the reliability of supergravity description and compare the
results against analysis on field theory side. The amount of
supersymmetry and available techniques should allow us to make
significant progress in probing these issues further. We plan to
report on these findings in future publications. 

\section*{Acknowledgements} AH thanks O.~Aharony, S.~Hirano, and P.~Ouyang for collaboration in \cite{Aharony:2009fc} on which much of this work was based. We also thank
P.~Argyres,
C.~Closset,
N.~Itzhaki,
Y.~Nomura,
I.~Yaakov, and
M.~Yamazaki for discussions.

\bibliography{sugra}\bibliographystyle{utphys}

\end{document}